\documentclass[aps,prl,twocolumn, comment, superscriptaddress,groupedaddress,amsmath,amssymb,floatfix,nofootinbib]{revtex4}  
\usepackage{amsmath,amssymb,bm,natbib}
\usepackage{epsfig}
\usepackage{graphicx}
\usepackage{slashed}
\newcommand{\beq}{\begin{equation}}
\newcommand{\eeq}{\end{equation}}
\newcommand{\bea}{\begin{eqnarray}}
\newcommand{\eea}{\end{eqnarray}}
\newcommand{\ba}{\begin{align}}
\newcommand{\ea}{\end{align}}
\newcommand{\bfig}{\begin{figure}}
\newcommand{\efig}{\end{figure}}

\newcommand{\D}{\displaystyle}

\newcommand{\gev}{\, \text{GeV}}
\newcommand{\fm}{\, \text{fm}}

\newcommand{\tin}{t_{\text{in}}}

\usepackage{color}

\newcommand{\omnes}{{\cal{O}}}

\begin{document}
\begin{flushright}
\end{flushright}
\phantom{}
\vspace*{-17mm}
\title{Electromagnetic charge radius of the pion at high precision}

\author{B.Ananthanarayan}
\affiliation{Centre for High Energy Physics,
Indian Institute of Science, Bangalore 560 012, India}
\author{Irinel Caprini}
\affiliation{Horia Hulubei National Institute for Physics and Nuclear Engineering,
 077125 Bucharest-Magurele, Romania}
\author{ Diganta Das}
\affiliation{Physical Research Laboratory,  Navrangpura, Ahmedabad 380 009, India}
\affiliation{Department of Physics and Astrophysics, University of Delhi, Delhi 110007, India}

\begin{abstract}
We present a determination of the pion charge radius from high precision data on the pion vector form factor from both timelike and spacelike regions, using  a novel formalism based on analyticity and unitarity. At low energies,  instead of the poorly known  modulus of the form factor, we use its phase,  known with high accuracy from
Roy equations for $\pi\pi$ elastic scattering via the Fermi-Watson theorem. We use also the values of the modulus at several higher timelike energies, where the data from $e^+e^-$-annihilation and $\tau$-decay are mutually consistent, as well as the most recent measurements at spacelike momenta. The experimental uncertainties are implemented by Monte-Carlo simulations. The results, which do not rely on a specific parametrization, are optimal for the given input information and do not depend on the unknown phase of the form factor above the first inelastic threshold.  Our prediction for the charge radius of the pion is 
$r_\pi=(0.657 \pm 0.003) \fm $, which amounts to an increase in precision by a factor of about 2.7 compared to the PDG average.

\end{abstract}

\maketitle
\emph{Introduction.---}
The electromagnetic charge radius of the pion is a fundamental observable of the strong interactions, with a long history dating from over half a century.   As with similar observables like the  proton radius or the pion-nucleon $\sigma$ term, its accurate determination is crucial for precision tests of the Standard Model at low energy, especially of Chiral Perturbation Theory (ChPT) and lattice QCD.  

In a relativistic theory, the mean charge radius squared is related to the slope at $t=0$  of the pion electromagnetic form factor  $F(t)$, according to the expansion  $F(t) = 1 + \frac{1}{6}\langle r_\pi^2 \rangle t + O(t^2)$. 
The most recent  value of the pion charge radius, $r_\pi\equiv\langle r_\pi^2 \rangle^{1/2}$, quoted by PDG \cite{PDG}
\begin{equation}\label{eq:PDG}
r_\pi=(0.672 \pm 0.008) \fm,
\end{equation}
 is obtained mainly from  values of the form factor measured at small spacelike momenta $t<0$ from $e\pi\to e\pi$ and $eN\to e \pi N$ processes, extrapolated to $t=0$ using  simple dipole-like parametrizations.  
ChPT at two-loop order \cite{Bijnens:1998fm}  in the timelike region, $t>0$,  yields the best value
 $\langle r_\pi^2 \rangle = (0.437\pm 0.016) \fm^2$, {\em i.e.}  $r_\pi=(0.661 \pm 0.015) \fm$. The calculations in lattice QCD, summarized in \cite{Aoki:2016frl}, are consistent with these values, but have not yet reached the same precision.

There is a large amount of information on the pion form factor which is not used in the radius extractions quoted above. For  $t>4m_\pi^2$, where $F(t)$ is a complex function, its modulus is measured either from  $e^+e^-\to \pi^+\pi^-$ annihilation or, using isospin symmetry, from the  $\tau^- \to \pi^-\pi^0\nu_{\tau}$ decay. This input involves however  energies more distant from $t=0$, so its use for the radius extraction requires a nontrivial analytic continuation.  The properties of the form factor that follow from causality and unitarity  play here an important role. It is known that $F(t)$ is analytic  in the complex $t$ plane with a unitarity cut along the  region $(4 m_\pi^2,\infty)$  of the real axis, and its phase below the first inelastic threshold, in the limit of exact isospin symmetry, is equal  by the Fermi-Watson theorem \cite{Fermi:2008zz,Watson} to the $P$-wave phase shift of  $\pi\pi$ elastic scattering.  
Many analyses  which exploit these properties 
have been performed over several decades, the pion form factor being actually one of the observables most investigated in  dispersion theory. However, the standard dispersive relations, written only in terms  of the phase (the so-called Omn\`es representation), or the modulus, or the imaginary part,  always require some poorly known or unavailable input. Therefore, although in some papers (for instance Refs. \cite{TrYn, Hanh2017}) the reported uncertainty is quite small, a certain model dependence is unavoidable in the standard dispersive calculations of the pion radius.

In this work we apply  an alternative mathematical formalism, proposed in \cite{IC} (see also \cite{Abbas:2010jc}),  which exploits in an optimal way analyticity, unitarity and the  information available on the form factor. Essentially, the formalism is a mixed phase-modulus dispersive representation, which uses as input at low energies the  phase,  very precisely known from $\pi\pi$ scattering, and at higher energies the modulus, measured by high precision experiments.  Using techniques of optimization theory for analytic functions, with no specific parametrization, we obtain for each input upper and lower bounds on the charge radius. The statistical distribution of the experimental input was then accounted for by Monte Carlo simulations, to convert the bounds into allowed intervals with definite confidence levels.
This work considerably improves our previous study Ref.~\cite{Ananthanarayan:2013dpa}, by properly treating the experimental errors and by including more experimental data. Similar techniques have been applied recently in Ref.~\cite{Ananthanarayan:2016mns} for evaluating the low-energy hadronic contribution to muon $g-2$.

\emph{Theoretical and experimental inputs.---}
  As shown in \cite{EiLu}, the first significant inelastic threshold for the pion vector form factor is produced by the 
$\omega\pi$ channel at $\sqrt{\tin}=m_\omega+m_\pi=0.917\,\gev$. 
 Below  $\tin$ we take the phase of the form factor as  the $P$-wave phase shift of $\pi\pi$ elastic scattering, which has been calculated very accurately from Roy equations and ChPT. We use the phase shift calculated in \cite{ACGL,Caprini:2011ky} and the so-called ``constrained'' fit to data from \cite{GarciaMartin:2011cn}, which we refer to as Bern  and Madrid phases, respectively.

 Above  the  inelastic  threshold, where the Fermi-Watson theorem is no longer valid and the phase  of the form factor is not known, we make use of the measurements of the modulus. In our conservative approach, we do not require the pointwise knowledge of $|F(t)|$, but adopt a weaker condition, expressed by a weighted integral of  $|F(t)|^2$. Several weighted integrals have been investigated in the previous work \cite{Ananthanarayan:2013dpa}, and stability of the results was proved for a large class of weights. Here we adopted the specific condition
\beq\label{eq:L2}
 \D\frac{1}{\pi} \int_{\tin}^{\infty} \frac{dt}{t} |F(t)|^2 \leq  I,
 \eeq
for which the available information allows an accurate and conservative estimate of the upper bound $I$. We evaluated the integral using the \emph{BABAR} data \cite{Babar} on the modulus $|F(t)|$ 
from  $\tin$ up to $\sqrt{t}=3\, \gev$, smoothly continued with  
a constant value for the modulus in the range $3\, \gev \leq \sqrt{t} \leq 20 \gev$,  and  a $1/t$ decreasing modulus at higher 
energies, as predicted by QCD scaling \cite{Farrar:1979aw,
Lepage:1979zb}. This led to the value  $ I=0.578 \pm 0.022$, 
where the uncertainty is  due to the \emph{BABAR} experimental errors. The contribution of the range above 3 GeV to the integral is actually of
only $1\%$. As discussed in previous works \cite{Ananthanarayan:2013dpa, Ananthanarayan:2016mns}, the assumed behaviour of the modulus above 3 GeV largely overestimate perturbative QCD calculations \cite{Melic:1998qr}, so the quoted value of $I$  is a conservative upper bound of the integral (\ref{eq:L2}). This leads to  weaker bounds on the charge radius, due to an exact monotony property of these bounds with respect to the value of $I$ \cite{Abbas:2010jc, Ananthanarayan:2013dpa}, which makes our  predictions conservative. For the same reason, in the numerical analysis
we have used as input for $I$  the  quoted central value increased by the  error. 

The measurements of the modulus of $F(t)$  below the inelastic threshold are expected to further improve the precision.  This input will be useful only if one can identify an energy range,  sufficiently close to the point $t=0$ in order to play a role in the radius extraction,  and where  accurate measurements are available.  Unfortunately, at very low energies the data have still large errors and there are discrepancies between the experiments. As shown in \cite{Ananthanarayan:2013dpa, Ananthanarayan:2016mns},  the above requirements are  satisfied by the range $(0.65-0.71) \gev$. The number of experimental points in this range varies from experiment to experiment. For the $e^+e^-$-annihilation experiments, there are 2 points each from CMD2 \cite{CMD2} and SND \cite{SND}, 
26 points for \emph{BABAR} \cite{Babar, Babar1}, 8 points each for KLOE 2011 \cite{KLOE2} and KLOE 2013 \cite{KLOE3} and 10 points for BESIII  \cite{BESIII}. 
For the $\tau$-decay experiments, there are 3 points each for CLEO \cite{Anderson:1999ui}, ALEPH  \cite{Schael:2005am, Davier:2013sfa} and OPAL \cite{Ackerstaff:1998yj},
and 2 points for Belle  \cite{Fujikawa:2008ma}. We note that, compared to Ref. \cite{Ananthanarayan:2013dpa}, we include now in addition  the very recent data of BESIII   \cite{BESIII},  the alternative KLOE analysis \cite{KLOE2}, and the $\tau$-decay data from \cite{Anderson:1999ui}-\cite{Fujikawa:2008ma}.  Several corrections, discussed in detail in Appendix B of Ref. \cite{Ananthanarayan:2016mns}, have been  applied to the data  in order to obtain the proper values of  $|F(t)|$ in the isospin limit, required in the formalism.  

As we mentioned, there is also rich experimental information on the pion form factor in the spacelike region, $t<0$. In our analysis we  used only  the values of the form factor measured   recently with high-precision by Jefferson Lab $F_\pi$ Collaboration \cite{Huber:2008id} at larger spacelike  energies, $F(-1.60\,\gev^2)= 0.243 \pm  0.012_{-0.008}^{+0.019}$ and
$F(-2.45\, \gev^2)=  0.167 \pm 0.010_{-0.007}^{+0.013}$. It is important to emphasize that our phenomenological input is complementary to that used in the determinations quoted by PDG. Finally, the  condition  $F(0)=1$ imposed by gauge invariance
is implemented in an exact way.

\emph{Calculation of $\langle r_\pi^2\rangle$ and its uncertainty.---} 
 We assume first that only one spacelike value $F(t_a)$ at a point $t_a<0$, and one value of the modulus $|F(t_b)|$ at a point $t_b$ in the specified range of the elastic region $(4 m_\pi^2, \tin)$ are known.  From the input described above, it is possible to derive an exact range for the first derivative of $F(t)$ at $t=0$, equal by definition to  $\langle r_\pi^2 \rangle/6$. 
We do not give the proof, which can be found in \cite{IC, Abbas:2010jc}, but simply quote the result. In order to proceed, we must introduce some notation. First make a change of variable from $t$ to  $z=\tilde z(t)$,  where
 \beq\label{eq:ztin}
 \tilde z(t) = \frac{\sqrt{\tin} - \sqrt {\tin -t}} {\sqrt{\tin} + \sqrt {\tin -t}},
 \eeq
and  define a new function $g(z)$ by
\beq\label{eq:gF}
 g(z) =F(\tilde t(z))\,[\omnes(\tilde t(z)) ]^{-1}  w(z)\, \omega(z) \,.
\eeq 
Here $\tilde t(z)$ is the inverse function of $\tilde z(t)$, $\omnes(t)$ is the Omn\`es function 
\beq	\label{eq:omnes}
 \omnes(t) = \exp \left(\D\frac {t} {\pi} \int^{\infty}_{4 m_\pi^2} dt' 
\D\frac{\delta (t^\prime)} {t^\prime (t^\prime -t)}\right),
\eeq
where $\delta(t)$ is equal to the $\pi\pi$ $P$-wave phase-shift $\delta_1^1(t)$  for
$t\le \tin$ and is an arbitrary smooth (Lipschitz continuous) function above $\tin$, 
and $w(z)$ and $ \omega(z)$ are two auxiliary functions, defined as
\beq\label{eq:outerfinal0}
w(z)=\sqrt{\frac{1-z}{1+z}},
\eeq 
\beq\label{eq:omega}
 \omega(z) =  \exp \left(\D\frac {\sqrt {\tin - \tilde t(z)}} {\pi} \int\limits^{\infty}_{\tin}  \D\frac {\ln |\omnes(t^\prime)|\, {\rm d}t^\prime}
 {\sqrt {t^\prime - \tin} (t^\prime -\tilde t(z))} \right).
\eeq 
Then one can prove the following exact inequality, expressed as positivity of the determinant:
\beq\label{eq:det}
\det\left|
\begin{array}{c c c}\vspace{0.2cm}
I-g(0)^2-g'(0)^2 & \bar g(z_a)  & \bar g(z_b)\\\vspace{0.2cm}	
	\bar g(z_a)  & \D \frac{z^{4}_{a}}{1-z^{2}_a} & \D
\frac{(z_a z_b)^2}{1-z_a z_b} \\
	\bar g(z_b)  &\D \frac{(z_a z_b)^{2}}{1-z_a z_b} & 
\D \frac{(z_b)^{4}}{1-z_b^2}\\
	\end{array}\right| \ge 0,
\eeq
where $z_a=\tilde z(t_a)$, $z_b=\tilde z(t_b)$ and $\bar g(z) =g(z) -g(0)-g'(0) z$. 

\begin{figure*}[htb]
	\begin{center}
\hspace{0.02cm}
\hbox{\hspace{0.03cm}
\hbox{
\includegraphics[scale=0.6]{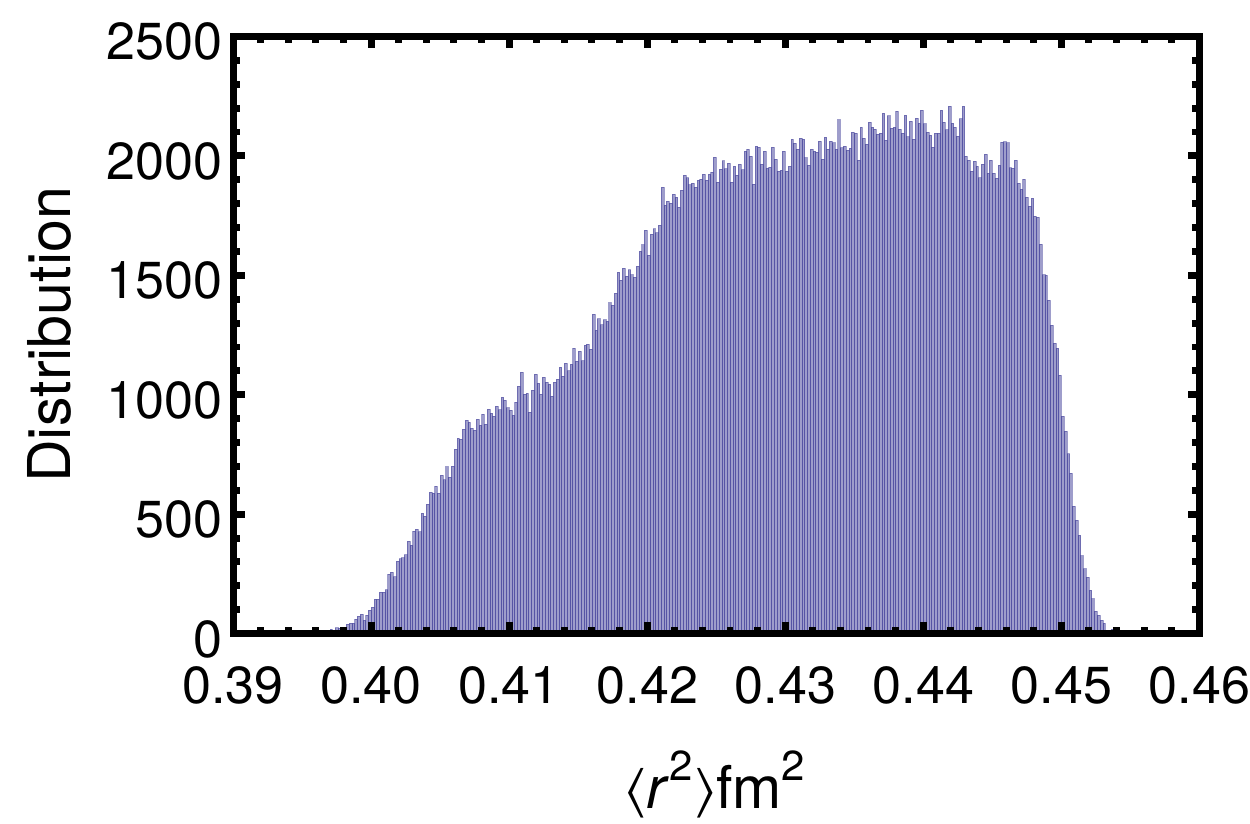}
\includegraphics[scale=0.6]{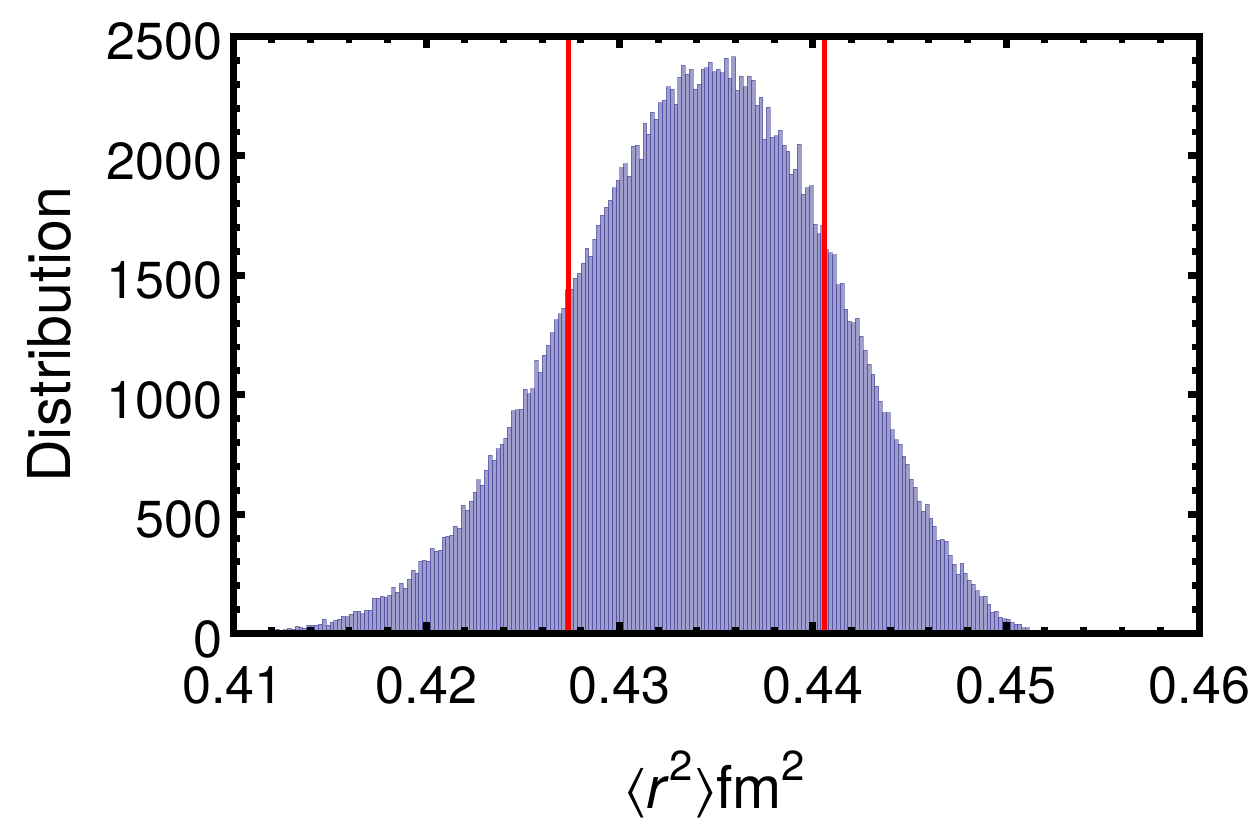} }}
\caption{Statistical distributions of $\langle r^2_\pi\rangle$ obtained using the Bern phase and one modulus measured by \emph{BABAR} experiment \cite{Babar}.   The left panel is obtained without spacelike input, the right one with input from  $t=-1.60 \gev^2$. The vertical lines correspond to 68.3\% CL.\label{fig:noSL}}
\end{center}
\end{figure*}

 As shown in \cite{Abbas:2010jc}, the dependence of the Omn\`es function $\omnes(t)$ on the arbitrary phase $\delta(t)$ above $\tin$ is exactly compensated in the product (\ref{eq:gF}) by the corresponding dependence of $\omega(z)$. Therefore,  except $\langle r_\pi^2 \rangle $, all the  quantities entering the determinant (\ref{eq:det}) are calculable and depend only on the given input.
From  (\ref{eq:gF}) it follows that $\langle r_\pi^2 \rangle $ enters only the expression of the derivative $g'(0)$, which can be written as $g'(0)=\xi +\eta \, \langle r_\pi^2 \rangle $, where $\xi$ and $\eta$ are real numbers. This implies that the determinant is a quadratic concave expression of $\langle r_\pi^2 \rangle$ and the positivity condition (\ref{eq:det}) can be written as 
\beq\label{eq:ABC}
A \,\langle r_\pi^2 \rangle^2 +2 B\, \langle r_\pi^2 \rangle +C\ge 0,  \quad\quad A\le 0.
\eeq
This inequality leads to a definite allowed range for $\langle r_\pi^2 \rangle$ if $B^2-AC\ge 0$ and has no solution if $B^2-AC< 0$. The latter case occurs when the phenomenological input adopted is inconsistent with analyticity. 

The inequality (\ref{eq:det}) involves only one spacelike value $F(t_a)$ and one timelike modulus  $|F(t_b)|$, which have been assumed to be known. The optimization formalism  applied here allows actually the simultaneous inclusion of several spacelike and timelike points. The general case, treated in \cite{Abbas:2010jc}, is expressed by a more general determinant, but leads to a condition of the same form (\ref{eq:ABC}) for the charge radius.  However, the simultaneous inclusion of many points leads to a system over-constrained by analyticity, which is difficult to handle numerically. Therefore, the most suitable approach for phenomenological applications is that adopted here, in which only one spacelike and one timelike constraint are imposed at the same time, the  results from different inputs being subsequently combined by a suitable averaging procedure.

From the inequality (\ref{eq:ABC}), an allowed range for $\langle r_\pi^2 \rangle$ can be obtained for every set of inputs.   However, except the condition $F(0)=1$, which is exact,  the input quantities are known only with some errors.
 One of the key aspect
of our calculation is a proper statistical treatment of these errors. This is achieved by randomly sampling each of the
inputs with specific distributions: the phase of $F(t)$, which is the result of a theoretical calculation, is assumed to be
uniformly distributed, while for the spacelike and the timelike data, which are known from experimental measurements, we adopt Gaussian  distribution with the measured central value as mean and the quoted error (the biggest error for spacelike data where the errors are asymmetric) as standard deviation.

For each set of values of the input statistical sample, if they are compatible, we calculate upper and lower value on $\langle r_\pi^2 \rangle$ from (\ref{eq:ABC}). Since all the values between these extreme points are equally allowed, we uniformly generate values of $\langle r_\pi^2 \rangle$ in between the bounds.  For convenience, the minimal separation between the generated points was set at $10^{-3}\fm^{2}$ and  for  
intervals  smaller than this limit no intermediate points were created. In this way, for each input from a spacelike energy and one timelike point in the region $(0.65-0.71) \gev$, we obtain a large  sample of values of $\langle r_\pi^2\rangle$. The results were proved to be stable against the variation of the size of the random sample and the minimal separation  mentioned above.

In Fig. \ref{fig:noSL} we show  for illustration two distributions  of $\langle r_\pi^2 \rangle$, obtained  using the Bern phase and one timelike modulus taken from the \emph{BABAR} experiment \cite{Babar}. 
Similar results have been obtained with the Madrid phase and other data on modulus. 
The left panel shows the distribution obtained  without the inclusion of a
spacelike datum (this case is obtained from Eq. (\ref{eq:det}) by removing the second row and column of the determinant). 
The histogram is rather flat and far from a normal distribution, which means that the input does not allow the extraction of a precise value of  $\langle r_\pi^2\rangle$  and its error. Adding the input from the spacelike point $t=-1.60 \gev^2$, we have obtained the distribution  in the right panel, shown actually to be very close to a Gaussian.  The explanation is that, adding the spacelike information, the input is much more constrained and many points in the Monte Carlo generated input do not satisfy the compatibility condition discussed below Eq. (\ref{eq:ABC}). The consistent values of the input lead to a normal distribution, which allows the extraction of the mean value and the standard deviation for the parameter $\langle r^2_\pi\rangle$. 

\begin{figure*}[htb]
\begin{center}
\hspace{0.00cm}
\hbox{\hspace{0.00cm}
\hbox{
\includegraphics[scale=0.67]{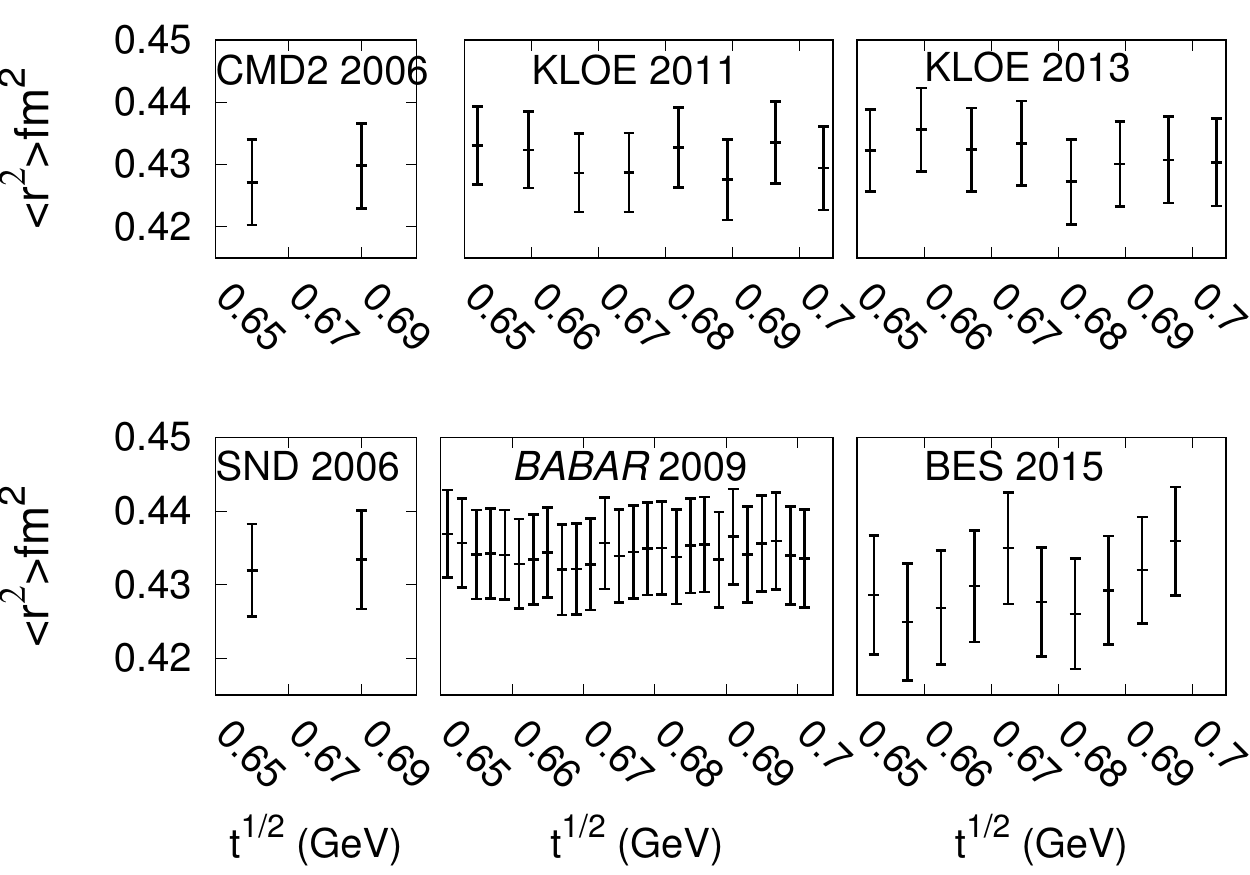}
\includegraphics[scale=0.65]{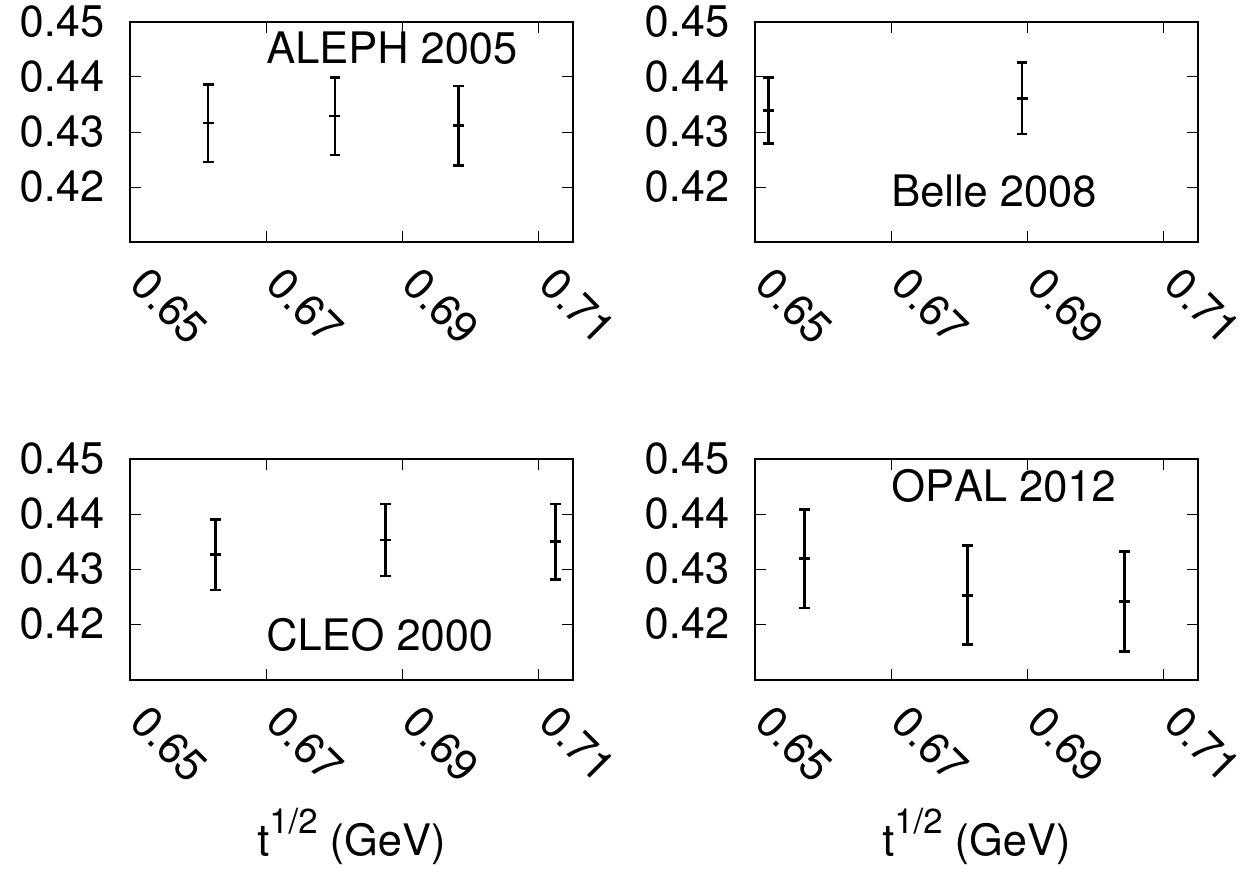}}}
\caption{68.3\% CL intervals of $\langle r_\pi^2\rangle$  for each input modulus in the region $(0.65 - 0.71) \gev$  measured in $e^+e^-$-annihilation and $\tau$-decay experiments, using the Bern phase and the spacelike datum at $t=-1.60 \gev^2$. \label{fig:fig2e}}
\end{center}
\end{figure*}

 In Fig.~\ref{fig:fig2e} we show the 68.3\% confidence limit (CL) intervals obtained in this way, for all the input data on modulus in the chosen range  $(0.65 - 0.71) \gev$,   for all experiments.  The Bern phase and the spacelike input at $t=-1.60 \gev^2$ have been used in this figure. Similar results are obtained with the Madrid phase, and also using the  spacelike datum at $t=-2.45 \gev^2$.

We take then the  average of the results obtained with input from various measurements.  Since the degrees of correlations between the measurements at different energies are expected to vary from one experiment to another,  we perform first the average of the values obtained with input from each experiment. The correlations are actually not known, therefore we apply the prescription proposed in 
\cite{Schmelling:1994pz} and adopted by PDG \cite{PDG}, where the effective correlation is determined from data themselves. Denoting by $v_i$ and $\sigma_i$ the central values and the standard deviations obtained from $n$ measurements, we define a generic covariance matrix $C(f)$ by $(C(f))_{ii}=\sigma_i^2$ and $(C(f))_{ij}= f  \sigma_i \sigma_j$ for $i\ne j$, where  $f\in (0,1)$ is an unknown global correlation.  Then the most robust average $\bar v$ is defined as standard  mean of $v_i$ weighted with the normalized weigths $\sigma_i^{-2}/\sum_j \sigma_j^{-2}$, and the 
standard deviation $\bar\sigma$  is  given by
\beq\label{eq:sigma}
\bar \sigma=\left [\sum_{i,j=1}^n (C(f)^{-1})_{ij}\right]^{\!\!-1/2},
\eeq 
where $f$ is the solution of the equation $\chi^2(f)/(n-1)=1$, for the standard definition of $\chi^2(f)$ 
\cite{Schmelling:1994pz}.

We have applied the above prescription to the values obtained from each $e^+e^-$-annihilation and $\tau$-decay  experiment, separately for the Bern and the  Madrid phase.  A conservative way of treating the information from spacelike $t$ has been adopted, by taking, for each timelike input, the simple average of the  central values and standard deviations obtained with each of the two considered  data.
It turned out that in all cases the ratio
$\chi^2(0)/(n-1)$ was less than 1 and increased for a positive
correlation, reaching unity for $f$ in the range $0.68 - 0.96$. 

\emph{Results.---} 
In Table \ref{table:rsq} we show the means and standard deviations obtained with the averaging prescription described above for the different timelike experiments.  
The results are mutually consistent among them, reflecting the consistency of the input data on modulus in the selected energy range.  The spread of the errors, ranging from the smallest (\emph{BABAR}) to the largest (OPAL), is actually not very large, which can be explained by the common information that entered the predictions.  We can therefore combine all the results, without the risk of a bias from a single experiment. 
Since the correlations between these values are difficult to estimate, we have applied again the averaging procedure described above
\cite{Schmelling:1994pz}, where an empirical overall correlation is extracted from the data. The correlation for all the ten values in Table \ref{table:rsq} was found to be 0.82 and 0.75 for the Bern and the Madrid phase, respectively, very close in both cases to the value $f$ where $\bar\sigma$ defined in (\ref{eq:sigma}) reached its maximum.  For a conservative estimate of $\bar\sigma$ we have adopted actually the maximum value of (\ref{eq:sigma}) for $f$ in the range (0,1).  This led to the predictions  
$0.4317 \pm 0.0044 \fm^2$ and $0.4323\pm 0.0039 \fm^2$  for $\langle r_\pi^2\rangle$,
for the Bern and the Madrid phase respectively.  Once again adopting a conservative treatment, we have combined the two determinations by taking their simple average  
$ \langle r_\pi^2\rangle = (0.4320 \pm 0.0041) \fm^2$, 
from which we obtained our final result  
 \begin{table}\vspace{0.3cm}
 \begin{tabular}{l c c }\hline \hline
 & ~~Bern phase & ~~~ Madrid phase\\\hline
 CMD2 06 & ~ $ 0.4281 \pm 0.0064$  & ~~ $ 0.4279 \pm  0.0061 $  \\
 SND 06 & ~ $ 0.4323 \pm 0.0059  $ &  ~~ $ 0.4327 \pm 0.0055 $  \\
 \emph{BABAR} 09  & ~ $ 0.4343 \pm 0.0048 $  & ~~ $0.4351  \pm 0.0046 $ \\
 KLOE 11 & ~ $ 0.4304 \pm 0.0055 $  & ~~ $ 0.4304 \pm 0.0048 $   \\
 KLOE 13 & ~ $  0.4311 \pm 0.0060 $ & ~~ $ 0.4313 \pm 0.0054 $  \\
 BESIII 15 & ~ $ 0.4293 \pm 0.0063 $ & ~~ $ 0.4319 \pm 0.0057 $  \\\hline
 CLEO 00 & ~ $0.4340 \pm 0.0060 $  & ~~ $ 0.4346 \pm 0.0054 $  \\
 ALEPH 05 & ~ $ 0.4315 \pm 0.0067 $  & ~~ $ 0.4318 \pm 0.0064 $  \\
 Belle 05 & ~ $  0.4347 \pm 0.0056 $  & ~~ $ 0.4356 \pm 0.0051 $ \\
 OPAL 12 & ~ $ 0.4266\pm 0.0082 $ & ~~ $ 0.4265 \pm 0.0079 $ \\
\hline\hline
 \end{tabular}\caption{Central values and standard deviations for the quantity $\langle r_\pi^2\rangle$
 obtained by the averaging prescription described in the text for each experiment. \label{table:rsq}}
 \end{table}
\beq\label{eq:result}
r_\pi=(0.657 \pm 0.003)\fm.
\eeq
The separate predictions obtained from $e^+e^-$ and $\tau$-decay data are
$(0.657 \pm 0.003)\fm$ and $(0.658 \pm 0.004)\fm$, respectively, showing that  (\ref{eq:result}) is dominated by the more precise $e^+e^-$ data.

The central value (\ref{eq:result}) is lower than the PDG average (\ref{eq:PDG}),  a feature that seems to be common to the  determinations based on analyticity \cite{TrYn, Hanh2017}. We note however that (\ref{eq:result}) is consistent with the determinations based only on elastic $e\pi$ scattering data, without including the more model-dependent determinations from electroproduction data at low spacelike $t$ \cite{PDG}. If not settled by experimental considerations, this difference is a challenge for precise lattice calculations, expected in the coming years.

The present work proves the strength of the general principles of analyticity and unitarity, enforced by suitable mathematical techniques and  supplemented by Monte Carlo simulations for error assessment.  Our prediction (\ref{eq:result}) is based on input complementary to that entering the PDG average  (\ref{eq:PDG}) and  has an uncertainty smaller by a factor of almost 2.7, making it it the most precise model-independent determination to date.

\begin{acknowledgements}
\emph{Acknowledgments.---} We would like to thank H. Leutwyler for suggesting this investigation. I.C. acknowledges support from the Ministry of Research and Innovation, Contract PN 16420101/2016. BA is partly supported by the MSIL Chair of the Division of Physical and Mathematical Sciences, Indian Institute of Science.
\end{acknowledgements}


\end{document}